# The dual footprint of artificial intelligence: environmental and social impacts across the globe

Paola Tubaro

Centre of Research in Economics and Statistics (CREST), CNRS-ENSAE-Polytechnic Institute of Paris, Palaiseau, France, paola.tubaro@cnrs.fr

## Abstract

This article introduces the concept of the 'dual footprint' as a heuristic device to capture the commonalities and interdependencies between the different impacts of artificial intelligence (AI) on the natural and social surroundings that supply resources for its production and use. Two in-depth case studies, each illustrating international flows of raw materials and of data work services, portray the AI industry as a value chain that spans national boundaries and perpetuates inherited global inequalities. The countries that drive AI development generate a massive demand for inputs and trigger social costs that, through the value chain, largely fall on more peripheral actors. The arrangements in place distribute the costs and benefits of AI unequally, resulting in unsustainable practices and preventing the upward mobility of more disadvantaged countries. The dual footprint grasps how the environmental and social dimensions of the dual footprint emanate from similar underlying socio-economic processes and geographical trajectories.

# Introduction

The spectacular rise of artificial intelligence (AI) in recent years sparks both aspirations and apprehensions about automation, which some fear may lead to job loss, yet others see potential for bolstering productivity in the workplace. Additionally, there is a distinctly positive outlook on utilizing AI to address climate change (Rolnick et al. 2022). However, multiple uncertainties cloud the potential long-term transformative impacts of AI, whereas current evidence is gloomier. The material infrastructure underpinning AI triggers accelerated extraction of raw materials and consumes hefty amounts of energy and water to sustain heavy computation in ever-bigger data centres (Crawford 2021). In parallel, it has become clear that AI development does generate additional demand for labour, but precarious and poorly remunerated. Myriad low-level workers perform essential, albeit unrecognized and underpaid data enrichment tasks in support of AI models (Muldoon et al. 2024), typically without any job security or social protection (ILO 2021).

This article lays out the concept of the 'dual footprint' as a heuristic device to capture the commonalities and interdependencies between these different impacts throughout the creation, production, and deployment of an AI solution. Originally devised as an ecological measure and later expanded to encompass various environmental and human dimensions (Matuštík & Kočí 2021), the concept denotes here the total impacts on the natural and social surroundings that supply the resources necessary for AI's production and use. It is an indicator of sustainability as initially intended, though in a modified sense. Evolving from the original idea of signalling whether an activity is approaching its biophysical limits (Costanza 2000), the dual footprint grasps the degree to which the AI industry is failing to ensure the maintenance of the social systems, economic structures, and environmental conditions necessary to maintain its production. To achieve broader coverage and reach previously uncharted territory, the proposed concept loses some of the accounting flavour of extant footprint measures, allowing for a more descriptive interpretation. It primarily serves as a mapping tool, linking impacts to specific locations and to the people and groups that inhabit them.

Beyond its debt to ecology, the notion of the dual footprint is interdisciplinary and draws on recent research that challenges idealized narratives of AI as the sole result of mathematics and code, or as the fancied machinic replacement of human brains. Studies from informatics to political economy bring to the fore the material substrate of AI and its reliance on global value chains that resemble those of more traditional products such as textiles and electronics. By integrating these diverse bodies of literature, this article examines the dual footprint of AI within the broader context of globalization and its long-standing trends of outsourcing and offshoring. Emphasis is on how utilization of natural and socio-economic resources in the AI industry thrives on existing cross-country disparities.

The argument is based on within-case and cross-case analyses of two in-depth case studies, each illustrating AI-induced cross-country flows of natural resources and data labour. The first involves Argentina as a supplier to the United States, while the second includes Madagascar and its primary export destinations: Japan and South Korea for raw materials, France for data work. These two cases portray the AI landscape as an asymmetric structure, where the countries that lead the AI race generate a massive demand for inputs (for hardware and/or software) and depend on imports of raw materials, components, and intermediate goods and services. Core AI producers trigger the footprint and therefore should bear responsibility for it, but the pressure on (natural and social) resources and the ensuing impacts occur predominantly elsewhere. Cross-country value chains shift the burden toward more peripheral players, obscuring the extent to which AI is material- and labour-intensive.

The analysis is primarily conceptual and combines multiple sources of evidence, from scientific publications to grey literature, the websites and the online repositories of governments, of specialist agencies like USGS (United States Geological Survey), and of international institutions like the World Bank. While the article does not directly analyse primary empirical data, several years of fieldwork experience on various aspects of the value chains of AI inevitably shape the viewpoints adopted here.

This drain of resources toward AI engenders adverse effects in more peripheral countries. Mining notoriously generates conflicts, and data work conditions are so poor that other segments of society – from local employers to workers' families and even informal-economy actors – must step in to cover part of the costs. The current arrangements thus fail to ensure their own sustainability over time. Additionally, the aspirations of these countries to leverage their participation to the AI value chain as a development opportunity, and to transition toward leading positions, remain unfulfilled.

The environmental and social dimensions of the dual footprint emanate from similar underlying processes. Value chain structures inherited from previous waves of globalisation, combined with political and macroeconomic hardship in many countries, make natural and social resources systematically and cheaply available to technology leaders. Legacy power dynamics shape the global AI industry and perpetuate inequalities in the distribution of the ensuing impacts.

## AI value chains and their impacts

Contemporary AI is largely based on machine learning systems, capable of inferring patterns from data, without every step being explicitly coded: it is often said that a model is trained rather than programmed. While this approach offers ever-greater adaptability to novel situations and supports multiple real-world applications, it requires massive amounts of data and huge computational power to process them. From the rise of so-called deep learning in the mid-2010s to present-day pre-trained general-purpose models with billions of parameters, appetite for data has increased exponentially and evolution of hardware has accelerated. The resulting business opportunities are drawing a diverse array of players, such as chip manufacturers, data service providers, and their suppliers across the globe. At this scale, AI is an economic force that redistributes resources and reshapes life conditions.

To see this, recent scholarship foregrounds the material foundations of AI production—that is, the design, development, and commercialization of so-called smart systems prior to their deployment in the economy. Crawford (2021) conceptualizes AI production as dependent on 'earth', labour, and data. Natural resources are essential for manufacturing and operating the computational infrastructure that processes data, while human work serves to curate and refine these data. Most specialists analyse these processes separately, and the next two subsections follow this tendency in examining them, before proposing the dual footprint as a step forward to explore the interplay between them. Nevertheless, both strands of research increasingly converge in seeing these processes as inter-organizational configurations that frequently transcend national borders, referring to them as 'AI supply chains' or 'AI value chains.' While scholars often use these two terms interchangeably, the remainder of this paper follows the recommendation of Attard-Frost and Widder (2025) in favour of the latter. Supply chains describe linear flows of material inputs, while value chains also consider how value is added at each step, encompass internal and external stakeholders, and include upstream and downstream interactions, allowing for a full life-cycle perspective.

*The environmental value chains of AI*

The environmental implications of the information and communications technology sector have become a focal point of both public discourse and academic inquiry. Within this context, the data centre has emerged as a key site of interdisciplinary analysis, emblematic of the digital world's material underpinnings and the convergence of its most advanced developments (Edwards et al. 2024). In particular, data centres are essential for AI production, offering cutting-edge hardware for high-performance processing, usually with multiple computing units working together and supported by extraordinary memory capacity. In anticipation of growing demand for AI-related services, investment in data centres is growing worldwide, and the size of the largest ones is increasing (IEA 2025).

Attempts to measure the impact of AI production in data centres originated within the machine learning community itself, causing a significant stir with their findings. In a path-breaking article, Emma Strubell and her coauthors (2019) showed that the training of a single large language model in a data centre may consume as much electricity, and emit as much greenhouse gases, as five cars throughout their lifetime. With more computing done in data centres, energy consumption and carbon emissions have grown over time (Schwartz et al. 2020). Data centres consume around 20% of the metered electricity supply in Ireland today, and 25% in the US state of Virginia (IEA 2025).

At the same time, many data centres are investing heavily to improve their energy efficiency and reduce their dependence on high-emission electricity sources. This consideration led a team of Google scientists to predict that 'the carbon footprint of machine learning training will plateau, then shrink' (Patterson et al. 2022). However, this hopeful conclusion rests on the narrow focus of these earlier analyses on model *training*. Carbon emissions almost double if one also counts the services that data centres offer before and after model training: the sheer provision of equipment consumes energy, not to mention the facilities that allow use of the AI solution when it is ready (known as 'inference' in computer science), which mainly corresponds to answering users' queries in the case of a language model (Luccioni et al. 2023). Likewise, producing AI can consume millions of litres of freshwater, withdrawn or consumed for electricity generation and for cooling servers (Li et al. 2025).

Another strand of research, more qualitative in nature and grounded in the social sciences and humanities, has highlighted the political relationships between data centres and their (natural and human) surroundings, though not always in the context of AI production. These studies have for example unveiled the problematic location of some data centres in dry areas despite their water needs (Hogan 2015), the entanglements of private and public interests in promoting their construction (Brodie 2020), and their (sometimes rapid) ruination when their profitability plummets (Velkova 2023).

But the data centre is not all. Part of computer science research has embraced Life Cycle Assessment (LCA, Ligozat et al. 2022), an environmental study methodology which extends the analysis of AI production up- and downstream of the data centre stage. This is harder to achieve. Downstream, the rare studies of end-of-life treatments of digital equipment use non-comparable methodologies (Ficher et al. 2025) and largely miss the growing amounts of e-waste mass that go undocumented and constitute over three quarters of the whole (Baldé et al. 2024). Upstream, one should count the manufacturing of all the devices used for data acquisition, training, and deployment, beyond the data centre infrastructure. Building this hardware requires not only energy but also components, like microprocessors and batteries, and consequently raw materials, especially metals and minerals. With data centres becoming progressively more energy-efficient, most emissions come from hardware manufacturing and raw material mining (Gupta et al. 2022). The methodology developed by Adrien Berthelot and co-authors (2024) gives a hint of the amounts involved. They analyse a generative-AI-as-a-service, that is, a model which provides new content from users' prompts and whose inference is accessible through a website interface. User equipment, networks, and web servers are as essential to the deployment of AI-as-a-service as powerful processors are to model training. The authors find that one year of service run requires minerals and metals equivalent to the production of 5659 smartphones, almost entirely due to user equipment and web servers.

These results suggest that the total pressure that AI puts on natural resources could continue to expand, even in the optimistic scenario of diminishing energy and water consumption in data centres. The energy transition may even reinforce this trend because low-emission sources are mineral-intensive: for example, an offshore wind plant needs nine times more minerals than a gas-fired plant of the same capacity (IEA 2021). Not only do these sources require a larger quantity, but also a wider diversity of minerals. The world demand for minerals may double by 2040 if current policies remain unchanged and even quadruple if countries engage on a stronger decarbonisation path (IEA 2021).

The social sciences and humanities have also extended their gaze beyond the data centre, harnessing the concept of the (value or supply) chain to capture the multiple actors that contribute to AI production at different stages. Lehuedé (2025) connects data centres and the extraction of minerals such as lithium because both are key components of digital technologies' value chains. On this basis, he identifies commonalities between the different communities impacted and their efforts of 'negotiating with AI's value chain' (p. 1762). Valdivia (2024) prefers the term supply chain to indicate the various steps that precede the production of AI solutions in data centres, from extraction of minerals to manufacturing of high-performance chips and wires. She sees the chain as a production process that is split across formally autonomous actors, although unequal distribution of power submits everyone to the same discipline. Thus, the communities that are negatively impacted do not necessarily coincide with those that enjoy the final AI solutions.

This literature is heterogeneous and dialogue between its different currents is limited. Nonetheless, its findings largely converge around the idea that AI exerts significant pressure on natural resources throughout both its production and deployment phases. Importantly, this research suggests that the environmental impacts of AI cannot be fully subsumed under the study of data centres, despite substantial overlaps. The diverse range of resource demands associated with AI—including energy, water, and raw materials such as minerals—sets it apart from adjacent sectors such as electronics, where the issue of minerals dominates, and digital technologies writ large, where energy is the main concern. More importantly, the rapid expansion of AI has intensified production processes that increasingly depend on environmental inputs. The chain framework proves especially useful in conceptualizing the interface between technological trajectories, the political-economic forces driving them, and the lived experiences of affected communities across different contexts and geographies.

*The human value chains of AI*

Another body of literature attends to the software side of AI production and reveals that there would be no data for machine learning without human labour. This is not limited to the work of high-level professionals such as engineers and computer scientists. To generate training data for speech recognition algorithms, for example, some workers are paid to record their voice, others to transcribe and label these recordings, indicating speaker's sentiment or characteristics like background noise (Tubaro & Casilli 2022). With these 'AI preparation' activities upstream of model training, two more human contributions are needed at the stage of inference, namely 'verification' and 'impersonation' (Tubaro et al. 2020). Verification is the check of output accuracy, for example 'red teaming', a practice of adversarial testing whereby workers probe generative AI solutions for the identification of harmful outputs (Zhang et al. 2024). Impersonation occurs when humans manually accompany, and sometimes replace, the functioning of AI algorithms that cannot deal with outlying cases or unexpected situations. Considering this, AI appears more labour-intensive than narratives of automation suggest, and increasingly so: business sources estimate that the market for these human-made services was USD 3.77 billion in 2024 and will reach USD 17.10 billion by 2030 (Grand View Research 2024).

The ILO casts doubt as to whether this form of labour meets its decent work standards (Berg et al. 2018). Formal employee status is uncommon, and most workers lack proper coverage of labour regulations and guarantees. The majority operate through digital platforms that consider them as independent providers and dispute the applicability of labour law. Like the platforms that allow clients to arrange a ride or order food, those that offer data services hire workers on demand, for one-off assignments with no durable commitments. Scholars and policymakers initially equated data work to a subset of 'uberised' platform labour and saw misclassification as the key matter of contention.

This perspective gradually evolved. Contrary to ride-hailing and delivery platforms that require work to be performed in specific locations, data enrichment activities can be done remotely. This apparently simple property has momentous implications insofar as it enables a planetary market connecting technology companies with contractors and providers across geographical distances. In this context, the earlier focus on classification appears too narrow to reveal the multiple production nodes in AI data work, the various actors and activities involved, and the ensuing spatial reconfigurations. New insights come from value chain analysis (Anwar 2025). First, this approach explains why conditions are bleak even when data work is procured through more conventional firm intermediaries in addition to digital platforms. Chains are the result of outsourcing, whereby AI producers avoid any responsibility to manage, supervise, and set wages for data workers, while still benefiting from their productive efforts. Regardless of their employment status, these external providers contribute to AI companies' profits but cannot make claims on those profits. They all operate in highly competitive environments that put downward pressure on their remunerations (Tubaro et al. 2025a). Second, value chains can be global, and data work is increasingly offshored to countries where labour costs are lower, like India (Chandhiramowuli et al. 2024), Kenya (Muldoon et al. 2025), and Venezuela (Posada 2022, 2024). Data work provision inherits structural features of these countries, like the prevalence of informal labour arrangements (Surie & Huws 2023). The labour flexibilization that outsourcing/offshoring practices promote reinforces informality and integrates it into the AI value chain. In this way, technology producers tap into 'labour pools hitherto almost inaccessible to wage labour' (Altenried 2020).

Of note, these insights have come from research both in computer science and in the social sciences, often through comparable methods and, sometimes, interdisciplinary collaboration. There is a broad consensus in this literature that AI production capitalizes on existing global disparities to secure cheaper albeit sufficiently qualified labour. In this regard, AI mirrors the digital services industry, which has a long tradition of outsourcing and offshoring activities such as data entry and commercial content moderation – that is, the classification and removal of inappropriate material from social media. AI has nevertheless transformed, and meddled with, these activities: for example, content moderation is now performed jointly by algorithms and humans, where the latter train the former, verify their performance, and intervene when they fall short. Crucially, the increasing use of a value chain lens in recent research reveals that the social consequences of data-related labour extend beyond narrowly defined employment status and surpass the geographic boundaries of the countries that lead AI development. The measures taken in some countries to regulate platform labour may thus prove inadequate to tackle the wider and more diverse effects of AI data work.

### *Duality*

These analyses have brought to light striking similarities between the depletion of mineral and material resources on the hardware side, and the social impacts of data work on the software side. Not only are both inputs to AI production, but both are also detected by scrutinizing the full-fledged value chains of AI, beyond the narrow stage of model training and the specific setting of data centres. Both give no sign of diminishing in size. Both are largely shifted to countries outside the narrow circle of those that drive the AI industry today. Thus, their impacts should be studied jointly as a by-product of longer value chains – a defining feature of today's globalization, already common in many industries and now accelerating under the effects of AI. Nevertheless, we still lack a unified framework to understand how this duality unfolds in practice. Specifically, what are the similarities, differences, or otherwise connections between the two? To move forwards, the remainder of this article draws on concepts from sustainability studies and frames the analysis in terms of a 'dual footprint' of AI.

# From the carbon footprint to the material/labour footprint of AI

The footprint saw the light in the 1990s as an accounting method, using a single numeraire to sum all the resources that a population must produce for its consumption and for assimilation of its waste (Rees 1992). Intuitive and easy to communicate, the concept became popular both within and outside academia. The original 'ecological footprint', measured in land units, later evolved into multiple indicators, based on different methodologies but all meant to provide synthetic quantitative measurements and routinely used to monitor progress toward sustainability (Matuštík & Kočí 2021). The carbon footprint, successful because of its direct link to climate change, is the total amount of emissions directly and indirectly caused by an activity or product and is usually expressed in mass units (kilograms) of carbon dioxide ($CO_2$) equivalent. Another common variant is the water footprint, computed as volume of water used for an activity. Both are used in the above-outlined literature that assesses the impacts of AI development in data centres.

However, a value chain approach that encompasses hardware production cannot be content with the carbon (and water) footprint concepts, which fail to capture exhaustion of mineral resources as another AI-induced pressure on the environment. A useful complement is the 'material footprint', defined as total mass of materials used for an economic process, especially its 'resource depletion footprint' version which weighs each element in the sum by its depletion potential to account for its relative scarcity (Fang & Heijungs 2014). Compared to the carbon footprint, the resource depletion footprint is not only an indicator of *pressure* on natural resources, but also of *impact,* seen as the tendency to resource exhaustion. Recent work that seeks to estimate the amount of material resources needed to run a generative AI-as-a-service (Berthelot et al. 2024, discussed above) is compatible with this approach, though it does not explicitly refer to it.

The material / resource depletion footprint concepts emphasize how the lengthening of value chains since the 1990s has spread the impacts of production across countries. For example, a car purchased in the United States generates a footprint that embodies the resources used through the procurement of raw materials in (say) Australia, of intermediate inputs in China, and of components in Germany (Wiedmann & Lenzen 2018). A notable finding is that industrialised countries have not managed to grow at a higher rate than their use of natural resources. As wealth expands, countries cut their domestic portion of extraction through imports of raw materials from overseas, essentially shifting the burden of natural resource consumption toward (mainly) lower-income countries (Wiedmann et al. 2015). These findings echo human geography insights that transferring environmental harm to poorer regions hides its full extent to decision-makers in wealthier areas (Parsons 2023).

The AI industry is no exception. Leader countries are the United States and China, followed by the United Kingdom, Japan, some European nations like Germany and France, and Canada (Chakravorti et al. 2023). The countries with most data centres are (in decreasing order) the United States, Germany, the United Kingdom, China, Canada, and France (Taylor 2024). While carbon emissions occur locally, the value chains that procure hardware components and raw materials span the planet. There is evidence of growth in the global demand for such inputs, particularly minerals now deemed 'critical' such as lithium, cobalt, nickel, copper, and so-called 'rare earths.' The result is a rise in the number of mining operations worldwide and higher production capacity of those already established (USGS 2025). Many of these operations are located far from core AI sites. For example, the Democratic Republic of Congo is the main source of mined cobalt, accounting for three quarters of world production.

If the material footprint concept aligns with extant evidence on the use of natural resources in AI production, it is also applicable to labour. Broadly speaking, human work enters production and can therefore be analysed as any other input. Inspired by Adam Smith's view that inequalities impose labour on others, Ali Alsamawi and co-authors (2014) propose a variant of the concept that they call the 'employment footprint' of a nation, as its domestic employment plus that occurring along its cross-country chains of provision. Richer countries that enjoy a lifestyle supported by workers elsewhere are more labour-intensive than local production conditions suggest. Moana Simas and her team (2014) go further and assess how global production chains channel 'bad labour', encompassing occupational health damage, vulnerable employment, gender inequality, share of unskilled work, child labour, and forced labour. They observe a net flow of bad labour from lower- to higher-income regions, whereby production in the former contributes to over half of the bad labour footprint caused by the affluence of the latter. Other studies broadly confirm these conclusions, though sometimes stretching the scope to include also collective bargaining, hazardous work, and (lack of) social security (Gómez-Paredes et al. 2015), sometimes narrowing it down to occupational safety and health (Alsamawi et al. 2017). The terminology varies too, referring at times to 'indecent labour' (García-Alaminos et al. 2020) as the opposite of the ILO's decent labour standards.

This body of literature takes a macro perspective, analysing countries rather than specific sectors such as AI data work. Nevertheless, it suggests mechanisms that may explain the above-presented evidence of problematic conditions due to income volatility, widespread informality, and paucity of health and safety support (ILO 2021). Overall, the diverse variants of the footprint concept offer key components for an integrated approach to the broad impacts of AI throughout its value chain, encompassing both natural and human resources, across geographical boundaries. Especially the material and the (bad) labour versions help to map essential inputs, to link them to people, groups and countries, and to assess their social cost. They can reveal whether the arrangements that currently sustain the AI industry (fail to) make it environmentally and socially sustainable. The remainder of this article develops this idea under the name of a 'dual footprint'.

To achieve this, some adaptations are necessary. AI and data work are not well defined in current statistical classifications, so that their amounts and cross-country flows are hard to quantify, and detailed input-output analyses like those typically used in material footprint research cannot be undertaken. Another barrier is that (typically macro-level) value chain data are of little use to address micro- or meso-level questions, like those that concern a specific sector. Finally, not all aspects of the existing 'bad' and 'indecent' labour definitions are applicable to data work where, for example, there is no evidence of forced labour to date. Hence, the remainder of this article moves away from the standard understanding of the footprint as an accounting method and adopts a more qualitative, descriptive, and at the same time comprehensive angle, to capture important dimensions of the phenomenon even without detailed measures. For labour, it starts from identifying the signals of indecent work already identified in the literature and on this basis, it seeks to gauge their broader consequences. The perspective is exploratory, prioritizing breadth of outlook over precision, aiming to bridge the gap until more refined classifications are available.

# Two case studies

A case study approach enables an all-inclusive, in-depth analysis. Unlike experimental designs, case research does not impose controlled conditions and is therefore less suited to testing causal hypotheses, but it sheds light on the role of the real-life contexts in which the facts of interest occur. It is helpful for disentangling multifaceted and dynamic phenomena, especially when prior knowledge is scarce, or when social processes evolve quickly (Yin 2017). The cases chosen here are the flows of minerals and data work services from Argentina to the United States on the one hand, and from Madagascar to Japan and South Korea (minerals) and to France (data work) on the other. To narrow down the analysis, the first case focuses on lithium, of which Argentina is the second-largest exporter to the United States (USGS 2025), and which is crucial to produce the lithium-ion batteries used in most devices. The other case foregrounds the Ambatovy project in Madagascar, where cobalt and nickel are mined together. It represents the largest-ever foreign investment in the country, and one of the biggest in sub-Saharan Africa, from a joint venture between a Japanese and a South Korean company. The dual exports from both Argentina and Madagascar enable a comparison of the economic and political processes involved in making these resources available to AI development, and of their up- and downstream impacts. Conversely, the institutional, historical, and macroeconomic differences between the two cases offer a glimpse into the variety of forms that the phenomenon takes.

### *Argentina – United States*

A leader of the global AI race under many respects, the United States hosted the first computations of the carbon and water footprints of AI (see above). It has a significant production of raw materials (with, among other things, almost 17% of the world resources of lithium and 4% of cobalt, USGS 2025) and has been a reservoir of data workers since the mid-2000s (Ross et al. 2010). Today, it constitutes the largest market for data work services, hosting many platforms and intermediaries (for example, Oneforma, Outlier, Remotasks), and most of their AI-producing clients, both academic and industrial.

Argentina is an upper-middle-income country with an emerging technology industry. Its human resources and know-how place it in the top 25 AI nations, though in the lower tier (Chakravorti et al. 2023). It has a growing mining sector, driven by lithium of which it is the fourth-largest producer behind Australia, Chile and China, and the largest holder with 20% of the world's resources, on a par with Bolivia (USGS 2025). Its other highlights like gold, silver and copper, are all needed in the hardware industry (BMI 2023). Argentina's regulatory framework is export-friendly and open to international private investment. Even if its lithium production is only one third of Chile's, it is steadily increasing its capacity. However, economic instability in recent years has discouraged the investments that would allow processing minerals locally and largely maintains the country in the role of raw material exporter (ECLAC 2023). In addition to the United States, the main destinations of Argentina's lithium exports are China, Japan, and South Korea, where equipment manufacturing is concentrated. This upstream position exposes Argentina to market fluctuations, notably as growing worldwide production, research into less lithium-intensive technologies, and advances in recycling are driving down lithium prices after a peak in 2022. Another drawback are the social conflicts that surround mining when there are unresolved issues around revenue distribution or environmental impacts. In 2022, there were 28 such conflicts in Argentina, fewer than in Mexico and Chile but more than in Brazil (ECLAC 2024, p. 254).

The recent economic crisis has made platform data work for foreign (mainly US-based) clients attractive to a qualified segment of Argentina's workforce. Platforms recruit internationally, and Argentineans see them as an opportunity to earn income in hard currency and leverage the parallel markets to secure advantageous exchange rates. But platform work is largely unprotected and pays little, thereby comparing unfavourably with Argentina's conventional jobs, mostly subject to historically strong labour legislation. Additionally, the financial knowledge needed to avoid the official exchange rate can be a barrier, requiring a mix of virtual wallets, digital dollar accounts, cryptocurrencies, and black markets. The preferred choice is in-between: maintaining a main (salaried or self-employed) position in the local economy and adding platform tasks to earn extra income in US dollars (Longo et al. 2024). Although this means dedicating a limited time to data work every day, turnover is lower in Argentina than in neighbouring countries, suggesting an enduring need for these earnings. Many data workers are among the country's higher strata: men in their twenties and thirties, University-educated, digitally and financially savvy, with a main job elsewhere, and connected to informal markets (Tubaro et al. 2025b).

It is now possible to return to the question of where impacts occur. Meeting US demand for lithium and data work requires Argentina's willingness to trade raw materials and a workforce in quest of hard currency. Impacts thus extend from the AI-leading country originating this demand to its provider. They are partly positive: development of the mining infrastructure, contribution to the development of AI, access to dollars. But there is also a negative side, in terms of the depletion of the natural resource under study and the emergence of social conflicts around mining. Additionally, the labour that fuels data services is 'bad' for two main reasons. First, because platform data work pays little and does not offer labour protections, workers can afford to practice it only as long as their local employers provide benefits and security. Platforms are free-riders (Schor et al. 2020) and transfer the burden from their US headquarters to the Argentinean economy. Second, data workers' recourse to informal financial transactions to manage their dollars can trigger hidden costs at local level if it escapes taxation, thereby failing to contribute to institutional development.

*Madagascar – Japan/South Korea – France*

Already mentioned as emerging buyers of South American lithium, Japan, South Korea, and other East Asian countries have important intermediate processing and device manufacturing industries that procure raw materials through a mix of international trade and foreign direct investment. Although they also have a buoyant AI industry at home, they are major component exporters and this second case focuses only on this role in hardware production, highlighting their purchases of nickel and cobalt from African countries like Madagascar. In 2023, they absorbed about two thirds of the latter's exports of nickel and almost half of cobalt according to UN COMTRADE data.

Madagascar is indeed rich in natural resources, including extensive deposits of not only nickel and cobalt (mostly from Ambatovy, one of the largest lateritic nickel mining projects worldwide) but also natural graphite, some rare earth elements like tantalum, and other minerals such as chromium and ilmenite. In 2021, its mining sector accounted for 4.8% of GDP, 1.07% of total government revenues and 32.2% of total exports (EITI 2022). The mining code of 2015, updated in 2023, facilitates investment inflows by offering favourable tax treatment and free access to foreign exchange. Export-friendly policies, including the establishment of free-trade zones, also facilitate trade in services and have made Madagascar the second-largest provider of computing services within French-speaking Africa, despite the informal nature of many firms. Data work has found a place among the set of services that local businesses were already offering, which started with telephone customer support services (call centres) and now also include, among others, remote sales and data entry.

The third vertex in this triangle is France, Madagascar's former colonial ruler, with which it maintains historical, linguistic, legal, and institutional proximities, and which constitutes the largest overseas client of its business services industry. France is a strong contributor to AI research, with a national strategy, a breeding ground for start-ups, and sizeable private and public investment. It hosts growing research on the environmental dimensions of AI (for example Fichet et al. 2025). It is at the forefront of data work, with platforms like Wirk/Yappers which mainly serve the national market, and IsaHit which matches (mainly) overseas workers with French clients. Today, demand is largely moving toward data work providers in Madagascar and other parts of francophone Africa.

Despite its place in AI value chains, Madagascar has one of the world's highest poverty rates, resulting from a long-term downward spiral and recurrent political crises that wiped out its rare spurts of economic growth, ending up worse than much of sub-Saharan Africa (Razafindrakoto et al. 2020). Mining constitutes an enclave economy hardly capable of creating jobs and largely disconnected from the local productive fabric. A result of this disconnection is lack of awareness of the scale of the country's mineral wealth among almost half of the population, a situation that likely contributes to maintaining the *status quo*, favouring resignation (Razafindrakoto et al. 2020, p. 179). The country's elites have more realistic perceptions but believe that foreign companies' market power prevents locals from benefiting. There are also concerns around the impact of mining on the island's unique fauna and flora. Madagascar joined the Extractive Industries Transparency Initiative (EITI) and committed to disclose sectoral data more systematically, improve environmental monitoring, increase contract transparency, and measure tax revenues (EITI 2022). However, progress so far is limited, with evidence of negative effects of mining on nearby communities, especially around Ambatovy (TI-MG 2024).

Regarding data work, use of international on-demand platforms is limited in Madagascar, where very expensive internet access hinders work-from-home. The main intermediaries are small local companies, chiefly formal and in some cases informal or semi-informal. Formal companies obtain large contracts for (mostly French) technology start-ups (Le Ludec et al. 2023) while informal ones sometimes arise around re-intermediation practices (whereby a worker gets tasks assigned and redistributes them to family and friends). They pay salaries and provide workers with office space and computing facilities.

Most workers are young and qualified: nine out of ten are under 35 years of age, three quarters have higher education degrees, and over two thirds are men (Tubaro et al. 2025a). They often see data work as their first step into the job market after graduation.

With a recognized employer, these jobs may seem to give some degree of income security. However, they are often temporary, tie remunerations to performance metrics at least partly, and pay little relative to the cost of living in the capital Antananarivo. Hence many workers, especially among the youngest, still live with their parents, while those who have children struggle to make ends meet. Career progressions are difficult, because French clients maintain control over production pipelines or delegate intermediate management positions to more experienced contractors in French-speaking Northern Africa. Therefore, Madagascan workers construe their current roles as temporary and hope to either migrate abroad afterward – which is difficult due to increasingly restrictive policies in higher-income countries – or to transition to better jobs nearby — which is also hard because the local labour market stagnates, and because limited capital availability hinders entrepreneurship.

Even without considering informality, this qualifies as bad labour. The conditions and remunerations of data work are much more constraining in Madagascar than in Argentina, as there is no direct access to foreign currency, no local job to fall upon for social protection, and little margin to build upon this experience to transition to a more rewarding career. Here, data work contractors free-ride on the families and local communities that bear the costs of supporting workers when pay is too low. These impacts combine with those of mining activities, which sharpen inequalities and disadvantage neighbouring populations. Hence, the burden shifts this time to a much more peripheral country that durable economic and institutional weaknesses keep upstream in the value chain.

## Discussion

The above analysis takes a step back from stricter interpretations of the footprint concept as an accounting method in favour of a bird's eye view. This more heuristic, descriptive perspective reveals who is impacted, by showing how pressure on resources, and related effects, are spread all along the AI value chain. On the hardware side, the carbon (and water) footprints of data centre functioning, model training and inference mainly occur in the countries that lead AI development – in the two examples used here, the United States and France. Despite signs that the carbon footprint of model training may be decreasing, the material / resource depletion footprint, ensuing from extraction of raw materials for use in hardware manufacturing, is growing in size. It is more widespread geographically, reaching Argentina from the United States and Madagascar from Japan or South Korea. Hopes to leverage AI to tackle environmental issues, or even just to reduce the carbon footprint by transitioning to cleaner energy sources, miscalculate costs and benefits if they do not encompass the full value chain. They may be shifting the burden across national boundaries in ways that prolong inherited patterns of domination and deepen cross-country economic inequalities.

On the software side, impacts spread across countries too. At least part of the supply of data work for the United States and France comes from areas of the world where labour costs are lower. For different reasons, countries as diverse as Argentina and Madagascar are reservoirs of highly educated workers that lack attractive alternative opportunities in their domestic economies. Recent evidence is that the share of data work that comes from middle- and lower-income countries is rising (ILO 2021), resulting in high labour inputs despite myths of inexorable AI-fuelled automation.

The footprint concept refers not only to the location, but also to the nature of these impacts. Regarding mining, Argentina and Madagascar make their natural resources available through export- and investor-friendly policies, but they have to strike a difficult balance between, on the one hand, appetite for capital and monetary inflows, and on the other, the protection needs of local ecosystems and communities. As discussed above, tensions around mining have arisen in both countries. In turn, data work qualifies as bad labour insofar as it imposes a toll on society. Lacking protection from platforms, Argentinean data workers rely on the guarantees that their other job in the local economy offers. On their side, underpaid Madagascan workers secure housing by living with their parents. Put differently, other parts of society cover the social costs of AI data work. Developing an idea proposed by Julian Posada (2022), this means that the currently observed conditions are unsustainable insofar as they introduce dependencies and burdens their make their own preservation over the long term impossible without significant contributions from third parties. By paying so little, the AI industry fails to ensure the conditions of its own production.

These considerations illuminate the 'dual' nature of the footprint. In the two case studies explored here, the same country (resp. Argentina and Madagascar) exports both mining products and data work services. Among importers, the United States is present in all links of the AI value chain, although its demand for some inputs exceeds its domestic supply, whereas other countries occupy a leading position only in a specific segment of the chain. Emphasis on France for software, and on East Asian countries for hardware, illustrates this possibility and at the same time exemplifies the length and convolution of these chains. Notwithstanding these differences, the general tendency is that both raw materials and data work outputs flow towards the countries that lead the worldwide AI race. Geographical distance obfuscates these contributions and their impacts, making it harder for leading countries to recognize, and take responsibility for, the footprints that they generate overseas.

## Conclusions

The above analysis spotlights ways in which footprint concepts enrich the social sciences. They could especially benefit the critical literature that leverages the notion of *extractivism*. From its Latin American origins, this construct designates practices of appropriation that date back to colonial times, when entire world regions were taken as mere sources of raw materials to be reaped. Some scholars have expanded it beyond its initial emphasis on natural resources, to designate an overarching disposition applicable to many commodities and sectors, including digital industries (Mezzadra & Neilson 2017). If Crawford qualifies AI as an '*extractive industry*' (2021, p. 15, author's emphasis), extraction is a generic metaphor difficult to operationalize. It is more helpful as an 'organizing concept' (Chagnon et al. 2022) that illuminates the profound social forces behind contemporary industrial production at a relatively abstract level, on a par with 'globalization'. The footprint concept helps to transform these ideas into tools suitable for mapping concrete practices and patterns, amenable to precise empirical description, and eventually measurable. The footprint arose within and for the study of the sustainability concept, diametrically opposed to extractivism as the latter denotes inherently destructive and non-reciprocal processes and practices.

Footprint concepts can advance research on extractivism in two main ways. They can precisely assign impacts to specific locations and segments of the value chain, recognising they entail distinct consequences. They also help to conceptualize the extraction of labour in digital environments, previously theorised mainly in terms of how technologies unlock surplus appropriation from activities outside of conventional salaried employment (Gago & Mezzadra 2017). Bad labour footprint ideas apply to any working arrangement insofar as they capture pressure on the social resources mobilized to compensate the effects of poor remunerations, and impacts on workers' livelihood, career prospects, and health – as the cases of Argentina and (especially) Madagascar demonstrate.

Likewise, fruitful dialogue can emerge between footprint notions and efforts to apply dependency theory to the global digital economy (Valente & Grohmann 2024). Dependency theory emphasizes how unequal exchange between leading technology producers and peripheral users hinders development (Fernández Franco et al. 2024). The cases examined here highlight one contemporary dependency-perpetuating mechanism: lower-level providers integrated into global AI value chains often remain too subordinate to make claims on the ensuing value, preventing positive spillovers locally, and even drawing resources from other sectors. Development under dependence would only be possible if AI producers fully covered the social costs of the human and natural resources they consume.

There is arguably a contribution in the opposite direction too, with the social and economic sciences enriching the nascent literature on the environmental costs of AI. The above analysis reveals the geographical spread of the footprint, whereby pressure on resources and ensuing impacts occur far from the sites that drive the development of technologies and their marketization. Doing so, it highlights the value of extending footprint studies throughout the full AI value chain, encompassing all contractors together with the lead actor, and counting both the software and hardware dimensions. These results rest on ideas, empirical evidence, and methods from the social and economic sciences. Studies of data work do not customarily use the footprint concept, yet they can contribute to its development by showing how similar socio-economic processes drive the absorption of natural and human resources into the AI value chain and generate negative local effects. Likewise, case study analysis is less technical than is customary in footprint research, but it is essential to highlight the offshore locations and the social dimensions of the impacts under study. In the perspective of environmental studies, the argument developed here can serve as a pilot, whose results outline directions for future investigation and invite further efforts toward quantification. More generally, it displays the importance of cross-disciplinary dialogue to gain insights into the multifaceted transformations that surround the rise of new technologies.

Accordingly, the contribution of the present study lies mainly in its effort to set the stage for an integrated analysis of the dual footprint of AI, in a way that facilitates dialogue and mutual enrichment between different literatures, while also operationalizing the ideas of more abstract social theories.

This is not to deny the limitations, with focus only on two sets of countries, use of heterogeneous sources, and lack of insights on the end of life of AI solutions.

Important implications and suggestions for action derive from this study. One concerns AI ethics, whose current formulations largely disregard the environmental costs of technology and the underpaid labour of data workers (Jobin et al. 2019). Reimagining AI ethics to account for the dual footprint of AI can raise awareness among stakeholders, strive toward a better balance of power, and at least in principle, offer better recognition to all contributing actors along the value chain.

Application of due diligence legislation could improve the monitoring of working conditions and environmental impacts. Laws that have recently come into force in France and Germany set obligations for lead firms to take responsibility for respect of labour legislation, human rights, and environmental protection all along their value chains, even beyond corporate boundaries and national borders. Initially intended for large multinationals in general and not for technology companies specifically, they may need some adaptation for sectoral specificities.

A final suggestion follows in the footsteps of the scholars who, inspired by the early measures of the carbon footprint of AI, invited computer scientists to report systematically the emissions of their models. Off-the-shelf software tools are now available to support these computations (Bannour et al. 2021). These reporting efforts could be at the same time extended and softened. Extended, by including disclosure of any other environmental impacts in addition to emissions, and of any use of data labour – including at least number and location of workers, instructions provided, and remuneration. Softened, by allowing a verbal description when it is impossible to calculate exact amounts. In this way, the reporting will be more comprehensive and applicable to all researchers, whether they develop new AI or use it as a tool – as is increasingly the case in multiple disciplines.

## Acknowledgements

This research benefited from funding from the ECOS-SUD France-Chile programme.

## Author biographic details

Paola Tubaro is research professor at the National Centre for Scientific Research (CNRS) and teaches at the National School of Economics and Statistics (ENSAE) in France. At the crossroads of economics and economic sociology, her current research explores the human labour underlying the digital economy, the ethical dimensions of big data and machine learning, as well as the impact of artificial intelligence on social inequalities, the spread of (dis)information, and the conduct of scientific research.